\documentclass{article}
\usepackage{spconf,amsmath,graphicx}
\usepackage{pifont}
\usepackage{multirow}

\title{Polyphonic Sound Event Detection Using Capsule Neural Network \\on Multi-Type-Multi-Scale Time-Frequency Representation}
\name{Wangkai Jin\textsuperscript{1}, Junyu Liu\textsuperscript{1}, Jianfeng Ren\textsuperscript{2},Xiangjun Peng\textsuperscript{3}\thanks{This work was supported in part by the National Natural Science Foundation of China under Grant 72071116, and in part by the Ningbo Municipal Bureau Science and Technology under Grants 2019B10026.}}
\address{\textsuperscript{1}User-Centric Computing Group, University of Nottingham Ningbo China \\ 
        \textsuperscript{2}School of Computer Science, University of Nottingham Ningbo China \\
        \textsuperscript{3}Department of Computer Science and Engineering, The Chinese University of Hong Kong}
\begin{document}
%
\maketitle
\begin{abstract}
The challenges of polyphonic sound event detection (PSED) stem from the detection of multiple overlapping events in a time series. Recent efforts exploit Deep Neural Networks (DNNs) on Time-Frequency Representations (TFRs) of audio clips as model inputs to mitigate such issues. However, existing solutions often rely on a single type of TFR, which causes under-utilization of input features. To this end, we propose a novel PSED framework, which incorporates Multi-Type-Multi-Scale TFRs. Our key insight is that: TFRs, which are of different types or in different scales, can reveal acoustics patterns in a complementary manner, so that the overlapped events can be best extracted by combining different TFRs. Moreover, our framework design applies a novel approach, to adaptively fuse different models and TFRs symbiotically. Hence, the overall performance can be significantly improved. We quantitatively examine the benefits of our framework by using Capsule Neural Networks, a state-of-the-art approach for PSED. The experimental results show that our method achieves a  reduction of 7\% in error rate compared with the state-of-the-art solutions on the TUT-SED 2016 dataset.

\end{abstract}

\section{Introduction}
\label{sec:intro}

Polyphonic Sound Event Detection (PSED) is widely applied in practice (e.g. wildlife monitoring~\cite{wildlife1,wildlife2} and acoustic surveillance~\cite{surveillance}). However, the existing approaches suffer from a high error rate. The root cause for this high error rate is the interference and overlaps between sound events in the timeline, and such issues are evidenced by the challenges to develop practical solutions for PSED~\cite{sed-survey}. Tackling such an obstacle is important for PSED since researchers are working towards effective artificial auditory systems aiming to automatically classify simultaneous sound events and recognize their corresponding onsets and offsets accordingly.

There are growing interests and efforts to use Deep Neural Networks (DNNs) to improve PSED performance, for automatic and accurate detection and localization. Convolutional Neural Networks (CNNs) are first applied to improve PSED tasks in~\cite{CNN1,CNN2,CNN3}, but suffer from the ineffectiveness of representing sequential features of sound events. Then, DNN models with strong sequential processing modules, are utilized to better abstract the temporal relationships of input sound events, such as Recurrent Neural Networks (RNNs)~\cite{RNN2} and Convolutional Recurrent Neural Networks (CRNNs)~\cite{CRNN1,CRNN2,CRNN3}. Though these approaches can reflect the features properly, it's not sufficient to recognize highly-dynamic and complex patterns in PSED. More recently, Capsule Neural Networks (CapsNets), as a new DNN architecture, are introduced to handle dynamic and complex patterns. CapsNet can process much richer features, and enable multi-level feature interactions by dynamic routing between neurons (i.e. capsules). The attractive characteristics of CapsNet breed early attempts for Bird Sound Classification~\cite{caps-bird} and Weakly-labeled Sound Event Detection~\cite{early-caps-sed}. All prior works focus on exploiting different DNN models with the support of a single Time-Frequency Representation (TFR), and none of them can learn the discriminative yet complementary features represented in different TFRs.

Our goal is to maximize the exploitation of  the key acoustic patterns in different types and scales of TFRs. The key observation is that, different types and scales of TFRs can reflect different symptoms of overlapped/interfered events, which can be further exploited to mitigate the issues of overlapped/interfered events. To this end, our key idea is to symbiotically combine various types of TFRs and the same type of TFR of different scales for PSED models. We denote this methodology as "Multi-Type-Multi-Scale" TFR. We develop a framework to improve the performance of PSED, via a novel approach to adaptively fusing different models and TFRs symbiotically. Therefore, the impacts of overlapped/interfered sound events can be greatly mitigated, which substantially improves the overall PSED performance. 

\begin{figure*}
    \centering
    \includegraphics[width=.95\linewidth]{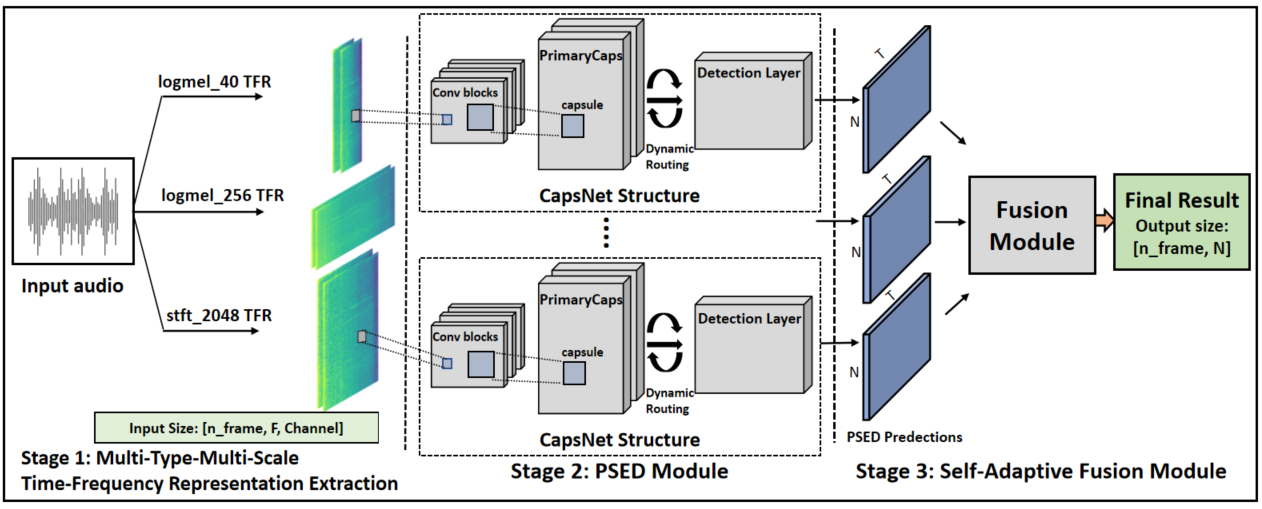}
    \caption{An overview of our proposed framework. Dimensions of the input vectors are illustrated in Sec~\ref{sec:TFR}. For PSED prediction results,$T$=256 and $N$ equals to the number of sound event types. }
    \label{fig:overview}
\end{figure*}

We quantitatively evaluate the effectiveness of our framework on CapsNet-enabled SED framework, the state-of-the-art approach for SED tasks~\cite{cap-sed}. Our evaluations on the TUT-SED 2016 dataset, shows a  reduction 7\% of segment-based Error Rate, compared with the original approach. We also perform an ablation study to understand the source of our improvements. Our results suggest that, our framework can effectively mitigate the overlapped/interfered sound events, which substantially increases the overall PSED accuracy.

\section{Proposed Method}

\subsection{Overview of Proposed Method}
\label{sec:overview}
Figure~\ref{fig:overview} shows an overview of our method. It has three stages: 1) \textit{Multi-Type-Multi-Scale Time-Frequency Representation Extraction}; 2) \textit{PSED Module}; and 3) \textit{Self-Adaptive Fusion Module}. In the first stage, various types of TFRs and multi-scale TFRs are extracted from the sound clips. Then the Multi-Type-Multi-Scale TFRs are utilized by separate PSED modules for detection. Finally a Self-Adaptive Fusion Module is used to fully leverage the predictions from different TFRs. In this work, we choose the CapsNet as the main model architecture as a proof-of-concept. 

\subsection{Multi-Type-Multi-Scale TFR Extraction}
\label{sec:TFR}

\begin{figure}[!h]

\begin{minipage}[b]{0.48\linewidth}
  \centering
  \centerline{\includegraphics[width=\linewidth]{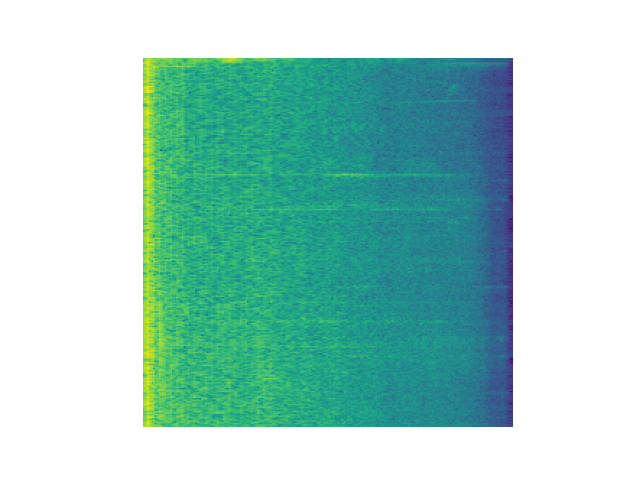}}
  \centerline{(a) logmel, 256 filters}
\end{minipage}
\begin{minipage}[b]{.48\linewidth}
  \centering
  \centerline{\includegraphics[width=\linewidth]{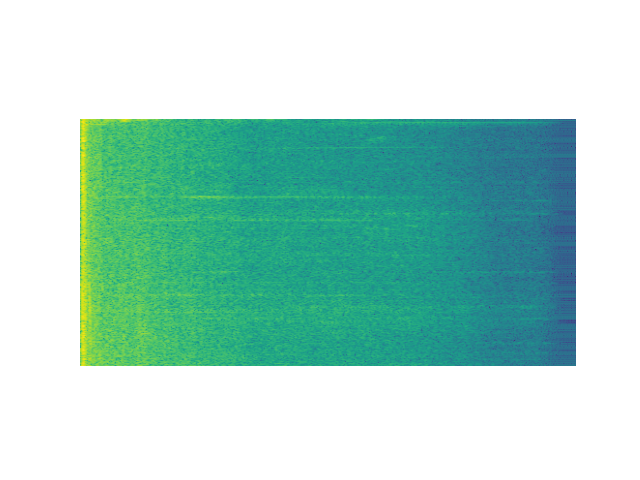}}
  \centerline{(b) STFT, 1024 n\_fft units}
\end{minipage}
\caption{(a) is the log mel spectrogram which apply a mel filter bank consisting of 256 triangular filters on a magnitude spectrum. (b) is a spectrogram transoformed via the Short-Time Fourier Transform with a window size of 1024. }
\label{fig:comp}
\end{figure}

 The first stage of our method is motivated by the fact that: different classes of TFRs can provide different patterns from the same sound clips, as they profile the characteristics of the audio clips in different ways, which can produce quite different representations, even merely from visual inspections (as shown in Figure~\ref{fig:comp}). Moreover, the same TFRs with different scales also exhibit significantly different characteristics, as the time-frequency trade-off in the TFR may lead to different encoding in terms of resolutions in time and frequency.

Our framework aims to symbiotically combine various types/scales of TFRs, and achieve comparable performance of different TFRs in a case-specific and highly-customized manner~\cite{huzaifah2017comparison}. We consider two representative acoustic spectral representations for the Multi-Type-Multi-Scale TFRs: 1) the magnitude characteristics obtained from the Short Time Fourier Transform (STFT); and 2) the coefficients of logarithmic Mel spectrogram, \textit{logMel}. All the extracted features from the audio files are binaural. This is because, as suggested by ~\cite{CRNN1,best-2016},  binaural features usually outperform monophonic spectral features for SED tasks. We configure the sample rate of STFT at 16kHz, and normalize the signals between -1 and 1 in input audios . For STFT, we set the frame size equal to 40 ms and the stride to be 20 ms. For each frame, 1204 points are used for computing STFT coefficients. Similarly, the logMel coefficients are extracted by applying a Mel Frequency Filter Bank with $n$ triangular filters on the extracted spectrogram via STFT, to map the frequencies into even-space mel scales and fit for the properties of the human ear. For the Multi-Scale TFRs for \textit{logmel}, we set $n$ to be \{40, 64, 128, 256, 512\}. As for the Multi-Scale TFRs for spectrograms, we set the length of STFT windowed signals $n_{FFT}$ to be \{1024, 2048\}. The number of selected scales for STFT is smaller as current settings can already construct high dimensional spectrograms, which is equal to $1+n_{FFT}/2$. Higher scales can incur unnecessary training cost and redundant features. For both spectral representations, we use a window with length of $T=256$ and binaural channels $C={1,2}$ to construct batches as the inputs for training and inference. Therefore, the input vector for spectrograms and \textit{logMel} coefficients is $X_{t:t+T-1} \in S^{256 \times F\times C}$, where F is  equal to the value of $n$ and ($1+n_{FFT}/2$) for each spectral representation respectively.

\subsection{PSED Module: Capsule Neural Network}
The second stage of our method uses Capsule Neural Network (CapsNet)~\cite{capsent-hinton} as the core module for PSED. We use CapsNet proposed in~\cite{cap-sed} for its rich feature representations and the potential to fully exploit the complex patterns in sound  clips. A CapsNet has three key components: 1) a series of CNN layers act as feature extractors for primary feature extractions and representations; 2) a Primary Capsule layer that transforms the extracted features into multi-dimensional units called \textit{capsule}, which represents the low-level features; 3) a Detection layer where the final results are determined by low-level capsules via a dynamic routing algorithm. In ~\cite{cap-sed}, the authors replace the densely connected layers in the original CapsNet for weights regularization with dropout modules and $L_2$ weight normalization. They also use a time-distributed layer in the output layer to apply weights at every time index.

\subsection{Self-Adaptive Fusion Module}

The final stage of our method is a module for adaptive fusion. This module adaptively aggregates the detection results from different types/scales of TFRs, as the final results. This module consists of two steps. First, the module obtains the Mean Square Error (MSE) for all results from different types/scales of TFRs on training data. Second, we propose a new formulation to aggregate the results from $m$ different TFRs to the final results, which is described as:
\begin{equation}
    \hat{y}=\frac{\sum_{k=1}^{m} w^{k}\left(\hat{y}^{k}-b^{k}\right)}{\sum_{k=1}^{m} w^{k}},
\end{equation}
where $m$ is the number of TFRs incorporated as model inputs, $w^{k}$ is the the reciprocal of the MSE value for each estimation result from each individual model, and $b^k$ is the local bias used for compensation before aggregation. The aggregated $\hat{y}$ is later activated by a dynamic threshold $\eta$ for each event, which marks the activation of the corresponding event if the estimation value exceeds the threshold. The values of $b^k$ and $\eta$ are jointly optimized during the fusion process, as the fusion process is completed block-by-block. The learned values for $b^k$, $\eta$ and $w^{k}$ are stored for later evaluation. 

\section{Experiments}
\label{sec:experiments}

\subsection{Data Set}
To evaluate our proposed method, we use the TUT Sound Events 2016  dataset~\cite{TUT-2016}, which is the official dataset used in  DCASE 2016 challenges (DCASE2016 Task3). The dataset was collected by researchers at Tampere University of Technology. It is categorized into two scenarios, which are ``Home" and ``Residential Area". The recordings in ``Home" scenario were captured in different indoor scenarios while recordings in the ``Residential Area" scenario were from various outdoor scenes. The total length for the recordings in the ``Home" and ``Residential Area" scenarios are 54 minutes and 59 minutes respectively. The recordings in these two scenarios are further partitioned into Development Set and Evaluation Set (i.e. approximately 70\%-30\% split).

\subsection{Evaluation Metric}


We use the segment-based Error Rate (ER) as the main evaluation metric, same as in DCASE Challenge 2016. We acknowledge that there are relatively new evaluation metrics in the recent DCASE challenges, such as Polyphonic Sound Detection Score (PSDS). However, we will still use the segment-based ER for a fair performance comparison with reported results using ER~\cite{cap-sed, best-2016, Adavanne2016, TUT-2016}. The segment-based ER is calculated by comparing the prediction results and the ground-truth in a one-second segment window, expressed as:
\begin{equation}
    ER =\frac{\sum S(s)+\sum D(s)+\sum I(s)}{\sum N(s)},
\end{equation}
where $s$ represents the segments in an audio clip, $S(s)$ stands for Substitution which is the number of False Negative (FN) and False Positive (FP) data points in the segment, $I(s)$ stands for Insertion which is the number of FPs after subtracting the substitutions, $D(s)$ stands for Deletion which is the number of FNs after subtracting the substitutions and $N(s)$ stands for the number of events in the ground-truth. 

\subsection{Experimental Settings}

\begin{table}[]
\caption{CapsNet Hyper-parameters in the TUT-SED 2016 dataset}
\begin{tabular}{ccc}
\hline
 & \begin{tabular}[c]{@{}c@{}}TUT 2016 \\ Home\end{tabular} & \begin{tabular}[c]{@{}c@{}}TUT 2016 \\ Residential Area\end{tabular} \\ \hline
\# of CNN kernels          & {[}32,32,8{]} & {[}4,16,32,4{]} \\ \hline
\# of CNN dim.             & 6 x 6         & 4 x 4           \\ \hline
\# of Pooling dim.         & {[}4,3,2{]}   & {[}2,2,2,2{]}   \\ \hline
\# of Primary Capsules     & 8             & 7               \\ \hline
\# of Primary Capsule dim. & 9             & 16              \\ \hline
\# of Output Capsule dim.  & 11            & 8               \\ \hline
\# of Routing Iterations   & 3             & 4               \\ \hline
\end{tabular}
\label{tab:hyper}
\end{table}

We report the hyperparameters of the best experimental results for the CapsNet in Table~\ref{tab:hyper}.  In total, we train the CapsNet in 100 epochs. The optimizer is AdaDelta optimizer, with an initial learning rate and decay rate of 1.0 and 0.95, respectively. Early-stop is leveraged to avoid over-fitting, which will halt the training process if the error rate stops to reduce for 20 epochs.

\subsection{Comparative Methods}
To verify the effectiveness of our method, we select the following methods for comparisons (denoted with their main models): 1) \textbf{RNN}~\cite{Adavanne2016}: Best solution in the DCASE 2016 Challenge, by using mel engery features; 2) \textbf{GNN}~\cite{TUT-2016}: Ranked the 2nd place in the DCASE 2016 Challenge, which uses mel energy; 3) \textbf{MLP}~\cite{best-2016}: Outperforms ~\cite{Adavanne2016} by using binaural log mel as features and  Multi-Layer Perceptrons  as classifiers; and 4) \textbf{CapsNet}~\cite{cap-sed}: State-of-the-art solution on the TUT-SED 2016 Dataset, which obtains the best results using binaural spectrograms.

\subsection{Experimental Results \& Ablation Study}
\label{sec:results}

\begin{table}[]
\caption{Segment-based ER on the TUT-SED 2016 Development and Evaluation Set. MFCC stands for the mel frequency cepstral coefficients. Logmel\_$n$ stands for log mel spectrogram filtered by $n$ mel-band filters while STFT\_$k$ stands for spectrograms transformed via STFT in  window size of $k$.}
\begin{tabular}{|c|c|c|c|c|}
\hline
          & \multicolumn{2}{c|}{Dev} & \multicolumn{2}{c|}{Eval} \\ \hline
Model     & Features        & ER     & Features        & ER      \\ \hline
RNN~\cite{Adavanne2016} & mel energy      & 0.91   & mel energy      & 0.81    \\
GNN~\cite{TUT-2016}   & MFCC            & 0.91   & MFCC            & 0.88    \\
MLP~\cite{best-2016}   & logmel\_40    & 0.78   & logmel\_40    & 0.79    \\
CapsNet~\cite{cap-sed}    & STFT\_1024      & 0.36   & STFT\_1024      & 0.69    \\ \hline
Our Method & \begin{tabular}[c]{@{}c@{}}logmel\_513 \\ +STFT\_2048\end{tabular} & \textbf{0.35} & \begin{tabular}[c]{@{}c@{}}logmel\_64 \\ +logmel\_128\end{tabular} & \textbf{0.62} \\ \hline
\end{tabular}
\label{tab:overall}
\end{table}

Table \ref{tab:overall} shows the segment-based ER on the TUT-SED 2016 Development set and Evaluation set. The ER of Development set and Evaluation set is the averaged result among two scenarios.   Our method achieves the best SED performance, by jointly leveraging logmel\_513 plus STFT\_2048 as acoustic features for Development set and logmel\_40 plus logmel\_128 for Evaluation set, with a reduction of 1\% and 7\%  ER on the Development Set and Evaluation Set respectively, comparing with the state-of-the-art~\cite{cap-sed}.  The overall ER on the TUT-SED 2016 dataset validates the effectiveness of our design.


\begin{table}[]
\centering
\caption{Segment-based ER on the TUT-SED 2016 Evaluation set. The underlined features and ER are reported in~\cite{cap-sed}.}
\begin{tabular}{|c|cc|}
\hline
Model                       & \multicolumn{1}{c|}{Features}                 & ER                        \\ \hline
\multirow{4}{*}{CapsNet}    & \multicolumn{1}{c|}{\underline{logmel\_40}}               & \underline{0.75}                      \\
                            & \multicolumn{1}{c|}{logmel\_64}               & 0.65                      \\
                            & \multicolumn{1}{c|}{logmel\_128}              & 0.78                      \\
                            & \multicolumn{1}{c|}{logmel\_256}              & \textbf{0.64}                      \\ \hline
\multirow{2}{*}{Our Method} & \multicolumn{1}{c|}{logmel\_64 + logmel\_128} & \textbf{0.62}                      \\
                            & \multicolumn{1}{c|}{logmel\_40 + logmel\_256} & 0.66                      \\ \hline
\multirow{2}{*}{CapsNet}    & \multicolumn{1}{c|}{\underline{STFT\_1024}}               & \underline{0.69}                      \\
                            & \multicolumn{1}{c|}{STFT\_2048}               & \textbf{0.66}                      \\ \hline
\multirow{2}{*}{Our Method} & \multicolumn{1}{c|}{logmel\_64 + STFT\_2048}  & \textbf{0.62}                      \\ 
                            & \multicolumn{1}{c|}{logmel\_513 + STFT\_2048} & 0.63 \\ \hline
\end{tabular}
\label{tab:breakdown}
\end{table}

To further understand the source of the improvements, we conduct an ablation study by breaking down the results into different features and discuss the key observations on the TUT-SED 2016 Evaluation set. We present the results trained on single CapsNet with single TFR and the results trained in our framework by using multiple TFRs, as shown in Table~\ref{tab:breakdown}. Note that we report results using at maximum 2 TFRs because 1) current fusion results already outperform the best results and 2) extensively training and fusing  on numerous TFRs is unpractical and cost-ineffective due to huge computational resource demands. From Table~\ref{tab:breakdown}, three key observations are concluded. First, \textbf{multi-scale TFRs of the same type can incur large variations in terms of ER}, as it is shown for logmel\_$n$ and STFT\_$k$ features in CapsNet, indicating the importance of  TFR scales.  Second, \textbf{the same TFR type with different scales can provide complementary features to improve SED accuracy}. By leveraging logmel\_64 and logmel\_128 in our framework, we achieves an ER of 0.62, which is less than the ER by using either of these two features solely as model inputs. Such phenomenon indicates that the acoustic features represented in different scales are beneficial for neural network inference, which also shows the feasibility of our self-adaptive fusion module to dig out complementary characteristics in outputs generated from different TFRs. Third, \textbf{different types of TFRs can provide complementary features to reduce SED ER}. In the table, we show the ER by leveraging logmel\_64 + STFT\_2048 jointly for SED. Similar to our second observation, the ER of joint-inference is less than the ER of single TFR for inference. This shows that different types of TFRs can also serve as complements for SED.

\section{Conclusions}
\label{sec:conclusion}

In this paper, we propose a novel framework for polyphonic SED tasks. We emphasizes the usage of Multi-Type-Multi-Scale TFRs for full exploration of the acoustic patterns in input audio clips. We validate our framework on CaspNet-enabled SED framework. which achieves a reduction of 7\%  in ER compared to the original approach.




\bibliographystyle{IEEEbib}
\bibliography{refs}

\end{document}